\documentclass[conference]{IEEEtran}

\IEEEoverridecommandlockouts
\usepackage{cite}
\usepackage{amsmath,amssymb,amsfonts}
\usepackage{algorithmic}
\usepackage{braket}
\usepackage{graphicx}
\usepackage{subcaption}
\usepackage{multirow}
\usepackage{textcomp}

\usepackage{xcolor}
\usepackage{natbib}
\usepackage{hyperref}

\def\BibTeX{{\rm B\kern-.05em{\sc i\kern-.025em b}\kern-.08em
    T\kern-.1667em\lower.7ex\hbox{E}\kern-.125emX}}

\makeatletter 
\newcommand{\linebreakand}{%
  \end{@IEEEauthorhalign}
  \hfill\mbox{}\par
  \mbox{}\hfill\begin{@IEEEauthorhalign}
}
\makeatother

\begin{document}

\title{Toward Quantum Enabled Solutions for Real-Time Currency Arbitrage in Financial Markets
}

\author{
\IEEEauthorblockN{Suman Kumar Roy}
\IEEEauthorblockA{\textit{TCS Research} \\
\textit{Tata Consultancy Services}\\
Kolkata, India \\
suman.r2@tcs.com}
\and
\IEEEauthorblockN{Rahul Rana}
\IEEEauthorblockA{\textit{TCS Research} \\
\textit{Tata Consultancy Services}\\
Kolkata, India \\
rahul.rana2@tcs.com}
\and 
\IEEEauthorblockN{M Girish Chandra}
\IEEEauthorblockA{\textit{TCS Research} \\
\textit{Tata Consultancy Services}\\
Bangalore, India \\
m.gchandra@tcs.com}
\and
\linebreakand

\IEEEauthorblockN{Nishant Kumar}
\IEEEauthorblockA{\textit{TCS BFSI} \\
\textit{Tata Consultancy Services}\\
Mumbai, India \\
nish.kumar@tcs.com}
\and
\IEEEauthorblockN{Manoj Nambiar}
\IEEEauthorblockA{\textit{TCS Research} \\
\textit{Tata Consultancy Services}\\
Mumbai, India \\
m.nambiar@tcs.com}

}
 
\maketitle
\renewcommand {\thefootnote} {\arabic{footnote} }
\setcounter {footnote} {1}
\def\footnoterule {\hrule width \columnwidth height 0.4pt \vspace{0.2cm}} 
\footnotetext {This paper has been accepted for poster presentation at the Symposium Celebrating the Quantum Century (SCQC) 2025 Conference.}
\begin{abstract}
Currency arbitrage leverages price discrepancies in currency exchange rates across different currency pairs to gain risk-free profits. It involves multiple trading, where short-lived price discrepancies require real-time, high-speed processing of vast solution space, posing challenges for classical computing. In this work, we formulate an enhanced mathematical model for the currency arbitrage problem by adding simple cycle preservation constraints, which guarantee trading cycle validity and eliminate redundant or infeasible substructures. To solve this model, we use and benchmark various solvers, including Quantum Annealing (QA), gate-based quantum approaches such as Variational Quantum Algorithm with Adaptive Cost Encoding (ACE), as well as classical solvers such as Gurobi and classical meta heuristics such as Tabu Search (TS). We propose a classical multi-bit swap post-processing  to improve the solution generated by ACE. Using real-world currency exchange data, we compare these methods in terms of both arbitrage profit and execution time, the two key performance metrics. Our results give insight into the current capabilities and limitations of quantum methods for real-time financial use cases.{\normalfont \textsuperscript{1}}

\end{abstract}


\section{Introduction}
In financial markets, currency arbitrage is the act of utilizing inconsistencies in exchange rates to make risk-free profits. This incorporates identifying cycles of currency conversions where the multiplication of exchange rates surpasses one. While the theoretical basis for arbitrage is thoroughly understood, executing profitable arbitrage strategies in practice remains difficult due to the speed at which opportunities appear and disappear in real-time trading scenarios. The need for fast, scalable, and accurate optimization methods makes this problem a strong case for quantum computing, which has the potential to solve combinatorial optimization problems more efficiently than classical approaches \cite{bib14}.

Several approaches have been proposed to model and solve the currency arbitrage optimization problem. Traditional methods rely on graph-based algorithms or linear/mixed-integer programming (MIP) techniques. For instance, arbitrage has often been modeled using negative logarithmic transformations to convert the problem into cycle detection in weighted graphs, as seen in early work by Bellman and Ford \cite{bib1}, and later applied in financial contexts. In paper \cite{bib2} the authors provide a mathematical framework for detecting and identifying triangular arbitrage opportunities. Linear Programming (LP) models, such as those used in portfolio and routing optimizations, have shown promise in capturing the profit-maximizing structure of arbitrage cycles, but they often struggle with execution time when scaled to realistic market scenarios \cite{bib3}. In paper \cite{bib4}, author presents two mathematical models for finding the optimal currency arbitrage opportunity, but none of the models includes the simple-cycle preservation constraints, which ensure no repeated currencies except at the start and end position of the arbitrage cycle.

With the advent of quantum computing, recent studies have explored its applicability to financial optimization. In paper \cite{bib5} a multi-objective portfolio optimization model is solved using quantum annealers. Gate-based approaches such as the Variational Quantum Eigensolver (VQE) have been extended to problems in finance, including portfolio optimization and asset allocation \cite{bib6}. Qubit efficient encoding scheme namely ACE for Variational Quantum Algorithms and circuit-level optimizations are also being explored to reduce the overhead associated with noisy intermediate-scale quantum (NISQ) devices \cite{bib7}. 

However, some work have addressed currency arbitrage specifically within a quantum framework with toy-size problems only \cite{bib4, bib8}, and to the best of our knowledge no prior work has benchmarked quantum techniques against strong classical baselines using real-world currency exchange data for currency arbitrage optimization problem.

In this work, we extend the currency arbitrage optimization model by introducing constraints that preserve simple-cycle within arbitrage opportunity. This addition ensures the feasibility and uniqueness of cycles identified by solvers, an important improvement over prior models \cite{bib4} that often allow redundant or invalid solutions. The proposed mathematical model employs a node-based formulation, enabling the incorporation of simple-cycle preservation (sub-tour elimination) constraints without introducing additional variables. In contrast, the edge-based formulation presented in \cite{bib8} does not account for this constraint, which may result in infeasible arbitrage cycles. We implement and evaluate this enhanced model using multiple quantum and classical optimization techniques.

Our key contributions are as follows:

\begin{itemize}
    \item Modeling: We propose an enhanced mathematical formulation for the currency arbitrage problem that includes simple-cycle preservation constraints to ensure valid arbitrage cycles.  

    \item Quantum and classical solving: We solve this model using a range of solvers- quantum annealing, gate-based quantum hybrid method (ACE) \cite{bib7}, and classical solvers (Gurobi and Tabu Search).

    \item Post-processing using classical local-search: We introduce multi bit swap as classical local search post-processing technique on the results from ACE to improve solution quality.
    
    \item Empirical evaluation: Using real-world currency exchange data, we benchmark all methods with respect to arbitrage profit and execution time, emphasizing execution latency as a critical factor for practical arbitrage.

\end{itemize}

Our findings shed light on the trade-offs between solution quality and computation time across different solving paradigms and offer insights about the potential of quantum computing for time-sensitive financial applications in the near future.

The rest of the paper is organized as follows: Section 2 presents the enhanced mathematical model, Section 3 details the quantum solution methods, Section 4 describes the experimental analysis and results, and Section 5 concludes with future research directions.

\section{Enhanced Mathematical Model}

Currency arbitrage (CA) can be modeled as a weighted directed graph, where vertices represent currencies and edges denote exchange rates, with edge weights defined as the negative logarithm of the exchange rates to convert the problem into finding negative cycles that indicate profitable arbitrage opportunities. Building upon this graph-based representation, we propose an enhanced mathematical model that extends existing formulations, such as those used in classical negative cycle detection \cite{bib4}, by introducing simple-cycle preservation constraints. These constraints ensure that only feasible and non-redundant trading cycles are considered, addressing a critical limitation in prior models where non simple-cycle (i.e., includes multiple repeated vertices) could lead to infeasible solutions. Our model formulates the arbitrage problem as a binary quadratic programming (BQP) problem, incorporating transaction costs and market constraints, and is designed to be compatible with both quantum and classical optimization techniques, enabling efficient exploration of profitable cycles while prioritizing execution time. The details of the BQP are as follows:

\hspace{-0.5cm}
\textbf{Parameters:}
\begin{itemize}
    \item[] $V$: Set of nodes (currencies); $E$: Set of edges in graph $G=(V, E)$.
    \item[] $w_{i,j}:$\; Currency exchange rate from currency $i$ to $j$, (e.g., $w_{\text{USD, EUR}}=0.8892$ in Table \ref{tab:input_data}).
    \item [] $N $: Number of currencies
    \item [] $K\geq 2(\in \mathbb{Z^+}) $:\;Cycle length, where $K \leq N$
\end{itemize}
\textbf{Decision variables:}
\begin{equation*}
  x_{i,k} =
    \begin{cases}
      1, & \text{if currency $i$ is in the $k^{th}$ position in the cycle}\\
      0, & \text{otherwise}
    \end{cases}
\end{equation*}
\textbf{Objective:} The objective is to find the most profitable cycle which is represented by:
\begin{equation}
\label{eq:1}
    min \;  \sum_{i, j \in V}\sum_{k=1}^{K} -log(w_{i,j})\cdot x_{i, k}x_{j, k+1}
\end{equation}

subject to:
\begin{equation}
\label{eq:2}
   \sum_{i=1}^{N} x_{i,k}= 1, \:\: \forall\; 1\leq k \leq K
\end{equation}
\begin{equation}
\label{eq:3}
\sum_{k=1}^{K} x_{i,k}\leq 1, \:\: \forall\; 1\leq i \leq N
\end{equation}
\begin{equation}
\label{eq:4}
    x_{i,k}x_{j,k+1} = 0, \:\: \forall\;(i,j)\notin E, \; 1\leq k \leq K
\end{equation}
\begin{equation}
\label{eq:5}
    x_{i,1} = x_{i,K+1},  \:\: \forall\; 1\leq i \leq N
\end{equation}

Equation (\ref{eq:2}) ensures that for each position of the cycle, exactly one currency is available, while Equation (\ref{eq:3}) guarantees that each currency is assigned to at most one position in the cycle. Together, Equations (\ref{eq:2}) and (\ref{eq:3}) effectively preserve simple-cycle. Additionally, Equations (\ref{eq:4}) and (\ref{eq:5}) ensure that non-convertible currencies are not placed in adjacent positions within the cycle and that the same currency is assigned to both the first and last positions to complete the cycle, respectively. In this mathematical model, the total number of binary variables are $N(K+1)$.

\section{Methodology}
To solve the CA problem, we used different approaches/solvers such as QA, ACE, TS, and Gurobi. The problem is initially formulated as a BQP problem as mentioned in the above section which is converted into a Quadratic Unconstrained Binary Optimization (QUBO) model for compatibility with QA, ACE and TS. In contrast, the Gurobi solver directly solves the BQP model.

\subsection{QUBO Formulation}
The corresponding QUBO of the BQP model is represented as:


\begin{multline}
\label{eq:6}
 min \Bigg( \Bigg. \sum_{i, j \in V}\sum_{k=1}^{K} -log(w_{i,j})\cdot x_{i, k} x_{j, k+1} +\\  \lambda_1 \sum_{k=1}^K (\sum_{i=1}^N x_{i,k}-1)^2 + \lambda_2 \sum_{i=1}^N \sum_{k=1,}^{K-1} \sum_{l=k+1}^K x_{i,k} x_{i,l} +\\ \lambda_3 \sum_{i=1}^N \sum_{\substack{j=1\\ (i, j) \notin E}}^N \sum_{k=1}^K x_{i,k} x_{j,k+1} + \lambda_4 \sum_{i=1}^N (x_{i,1}-x_{i,K+1})^2 \Bigg. \Bigg)
\end{multline}

where the second, third, fourth and fifth term of the above equation are the corresponding penalty terms of constraints eq.(\ref{eq:2}), eq.(\ref{eq:3}), eq.(\ref{eq:4}), eq.(\ref{eq:5}) respectively and $\lambda_1, \lambda_2, \lambda_3, \lambda_4$ are the respective penalty values. We have used  Verma and Lewis Method (VLM)\ \cite{bib8a} to find the optimal values of  $\lambda_1, \lambda_2, \lambda_3 \ \text{and}\ \lambda_4$.

\subsection{Quantum Annealing} 
Quantum computing utilizes quantum mechanical principles, such as superposition, entanglement and interference, to process information utilizing quantum bits (qubits). Within quantum computing, quantum annealers portray a specialized analog procedure separate from digital or gate-based systems. It is being designed explicitly for optimization problems which are extensively used in fields such as logistics, finance, materials science, and artificial intelligence. QA aim to efficiently identify the optimal solution from all possibilities. This approach includes converting an optimization problem into a Hamiltonian, where the system's lowest-energy state represents the optimal solution. Quantum annealing operates qubits through a preset annealing schedule, which initializes with a Hamiltonian which is easy-to-solve and changes slowly to problem Hamiltonian by using adiabatic theorem \cite{bib15}. Unlike classical simulated annealing, which depends on thermal fluctuations, QA is leveraging quantum tunneling to direct potential barriers, having advantages to avoid local minima in finding global optima. QA requires QUBO and it's equivalent Ising formulations for solving the optimization problem \cite{bib9}.

\subsection{Adaptive Cost Encoding} 
To tackle the challenge of solving large-scale QUBO problems on gate-based quantum computers, Adaptive Cost Encoding (ACE) algorithm \cite{bib7} leverages Variational Quantum Algorithms (VQAs) with a minimal encoding scheme. This approach exponentially reduces the qubit demands compared to traditional VQAs. For example, Variational Quantum Eigensolver (VQE) and Quantum Approximate Optimization Algorithm (QAOA), where required qubits are equal to number of binary variables in QUBO. In ACE to solve a QUBO problem with $n$ number of binary variables, the number of qubits reduce to $\left\lceil log_2(n)\right\rceil+1$. Here, $\left\lceil log_2(n)\right\rceil$ register qubits encode the state indices and an ancilla qubit represents the binary value corresponding to each state. The quantum circuit is prepared with a layer of Hadamard gates to create an equal superposition, subsequently a hardware-efficient ansatz is used to evolve the quantum state. Hardware-efficient ansatz includes sequence of parameterized single-qubit rotations and entangled gates. The quantum state has been provided by eq.(\ref{eq:7}) \cite{bib7}. 
\newline
\begin{equation}
\label{eq:7}
    \ket{\Psi(\theta)}=\sum_{i=1}^n c_i(\theta)\{\alpha_i(\theta)\ket{0}_a + \beta_i(\theta)\ket{1}_a\} \otimes \ket{\phi_i}_r
\end{equation}
where $\ket{\Psi(\theta)}$ defines the quantum state generated from the defined quantum circuit with set of parameters $\theta$ and the coefficients $c_i(\theta)$. The quantum states $\ket{\phi_i}_r$  are computational basis states of the register qubits. $\ket{0}_a, \ket{1}_a$ are quantum state and $\alpha_i$ and $\beta_i$ are respective amplitudes for the ancilla qubit. The probability of the $x_i$-th classical variable to get the value 1 or 0 is provided by $Pr(x_i = 1) = |\beta_i|^2$ and $Pr(x_i = 0) = |\alpha_i|^2$ respectively \cite{bib7}. During each iteration of the optimization, the quantum circuit is measured multiple times to obtain a probability distribution over the state indices. This distribution is decoded into a binary vector $x$ that represents a possible solution, which is used to optimize the cost function provided in eq.(\ref{eq:6}), where $x=[x_1, x_2, x_3, \dotsc, x_n]$.
\newline
\textbf{Local Search (LS):} To improve the solution quality, the binary vector $x$ goes through a classical post-processing step that includes a local bit-swap search. This approach sequentially estimates each bit of $x$ and computes the objective value along with checking the feasibility of the resulting bit string. If the flip enhance the objective value and ensures feasibility, the change is retained; otherwise, it is reverted. The local search required $\mathcal{O}(n)$ local bit flips, where each bit flip on vertex $i$ affects $\mathcal{O}(d_i)$ edges, with $d_i$ the degree of the vertex. Hence, the update of the objective value per bit flip is $\mathcal{O}(d_i)$ in each term of $x$ and the total complexity of the entire round is thus $\mathcal{O}(n^2)$.

\begin{figure}[!h]
\centering
\includegraphics[scale=0.4]{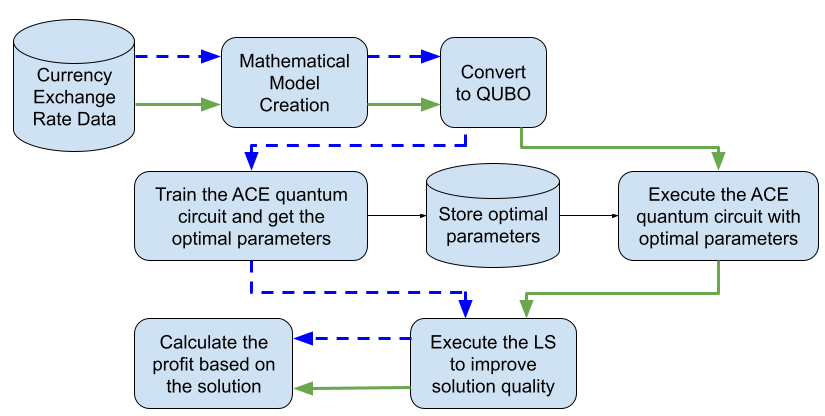}
\caption{ACE-LS Workflow for Currency Arbitrage}
\label{fig:ace_ls}
\end{figure}

Figure \ref{fig:ace_ls} shows the overall flow of the Currency Arbitrage, where the blue dotted line represents the training phase of the ACE circuit with training data to determine its optimal parameters, while the green solid line depicts the execution phase of the circuit utilizing these optimized parameters with execution data. In this figure, the currency exchange data has been taken from  \cite{bib13} to create the mathematical model and QUBO respectively. The idea is to train the ACE circuit and store the optimal parameters, then use those optimal parameters to run the circuit. A local search approach has been implemented to improve the generated solution quality.

\begin{table*}[ht]
\caption{Real-Time Bid and Ask Exchange Rates for 14 Currency Pairs}
\label{tab:input_data}
\begin{tabular}{|c|l|l|l|l|l|l|l|l|l|l|l|l|l|l|}
\hline
\multicolumn{1}{|l|}{} & \multicolumn{1}{c|}{\textbf{EUR}} & \multicolumn{1}{c|}{\textbf{USD}} & \multicolumn{1}{c|}{\textbf{GBP}} & \multicolumn{1}{c|}{\textbf{CAD}} & \multicolumn{1}{c|}{\textbf{CHF}} & \multicolumn{1}{c|}{\textbf{JPY}} & \multicolumn{1}{c|}{\textbf{AUD}} & \multicolumn{1}{c|}{\textbf{CZK}} & \multicolumn{1}{c|}{\textbf{HUF}} & \multicolumn{1}{c|}{\textbf{NZD}} & \multicolumn{1}{c|}{\textbf{SEK}} & \multicolumn{1}{c|}{\textbf{SGD}} & \multicolumn{1}{c|}{\textbf{DKK}} & \multicolumn{1}{c|}{\textbf{NOK}} \\ \hline
\textbf{EUR}           & 1                                 & 1.1245                            & 0.8412                            & 1.5694                            & 0.938                             & 162.844                           & 1.7445                            & 24.869                            & 401.986                           & 1.8989                            & 10.8786                           & 1.4559                            & 7.4594                            & 11.5808                           \\ \hline
\textbf{USD}           & 0.8892                            & 1                                 & 0.748                             & 1.3957                            & 0.8341                            & 144.806                           & 1.5512                            & 22.115                            & 357.492                           & 1.6892                            & 9.6761                            & 1.2947                            & 6.6336                            & 10.3004                           \\ \hline
\textbf{GBP}           & 1.1886                            & 1.3367                            & 1                                 & 1.8655                            & 1.1150                            & 193.575                           & 2.07385                           & 0                                 & 0                                 & 2.2575                            & 0                                 & 1.7305                            & 0                                 & 0                                 \\ \hline
\textbf{CAD}           & 0.6371                            & 0.7164                            & 0.5359                            & 1                                 & 0.5976                            & 103.757                           & 1.1115                            & 0                                 & 0                                 & 1.2100                            & 0                                 & 0.9275                            & 0                                 & 0                                 \\ \hline
\textbf{CHF}           & 1.0659                            & 1.1987                            & 0.8966                            & 1.6726                            & 1                                 & 173.576                           & 0                                 & 0                                 & 0                                 & 0                                 & 0                                 & 0                                 & 0                                 & 0                                 \\ \hline
\textbf{JPY}           & 0.0061                            & 0.0069                            & 0.0052                            & 0.0096                            & 0.0058                            & 1                                 & 0.0107                            & 0                                 & 0                                 & 0.0117                            & 0                                 & 0.0089                            & 0                                 & 0                                 \\ \hline
\textbf{AUD}           & 0.5731                            & 0.6445                            & 0.4821                            & 0.8994                            & 0                                 & 93.328                            & 1                                 & 0                                 & 0                                 & 1.0884                            & 0                                 & 0.8343                            & 0                                 & 0                                 \\ \hline
\textbf{CZK}           & 0.0402                            & 0.0452                            & 0                                 & 0                                 & 0                                 & 0                                 & 0                                 & 1                                 & 0                                 & 0                                 & 0                                 & 0                                 & 0                                 & 0                                 \\ \hline
\textbf{HUF}           & 0.0025                            & 0.0028                            & 0                                 & 0                                 & 0                                 & 0                                 & 0                                 & 0                                 & 1                                 & 0                                 & 0                                 & 0                                 & 0                                 & 0                                 \\ \hline
\textbf{NZD}           & 0.5264                            & 0.5918                            & 0.4428                            & 0.8261                            & 0                                 & 85.717                            & 0.9185                            & 0                                 & 0                                 & 1                                 & 0                                 & 0.7663                            & 0                                 & 0                                 \\ \hline
\textbf{SEK}           & 0.0919                            & 0.1033                            & 0                                 & 0                                 & 0                                 & 0                                 & 0                                 & 0                                 & 0                                 & 0                                 & 1                                 & 0                                 & 0                                 & 0                                 \\ \hline
\textbf{SGD}           & 0.6867                            & 0.7722                            & 0.5777                            & 1.0778                            & 0                                 & 111.842                           & 1.1982                            & 0                                 & 0                                 & 1.3044                            & 0                                 & 1                                 & 0                                 & 0                                 \\ \hline
\textbf{DKK}           & 0.134                             & 0.1507                            & 0                                 & 0                                 & 0                                 & 0                                 & 0                                 & 0                                 & 0                                 & 0                                 & 0                                 & 0                                 & 1                                 & 0                                 \\ \hline
\textbf{NOK}           & 0.0863                            & 0.097                             & 0                                 & 0                                 & 0                                 & 0                                 & 0                                 & 0                                 & 0                                 & 0                                 & 0                                 & 0                                 & 0                                 & 1                                 \\ \hline
\end{tabular}
\end{table*}


\section{Experimental Analysis}

\subsection{Experimental Setup}
DefaultQubit simulator from PennyLane \cite{bib11} is applied for ACE approach, as it supports reliable quantum circuit design with single and multi-qubit gates, while D-wave Ocean \cite{bib12} has been utilized for QA. The classical system supporting all these experiments consist of an 11th Gen Intel®Core™i5-1145G7 @ 2.60GHz×8 with 16GB RAM.
\subsection{Experimental Data}
The experimental analysis utilizes a real time currency exchange rate data \cite{bib13}. The details have been provided in Table \ref{tab:input_data} with exchange rates of 14 currencies for a particular timestamp. The diagonal elements are 1, indicating the exchange rate of a currency with itself. On the other hand, the off-diagonal elements provide the exchange rate between pairs (e.g., EUR/USD = bid price of EUR w.r.t USD = 1.1245, USD/EUR = ask price of EUR w.r.t USD = 0.8892). The exchange rates which show 0 indicate there is no direct exchange between those currencies. 

\subsection{Results}
We have experimented with different solvers which includes QA; ACE-LS, a gate-based quantum method; TS, a meta-heuristic solver \cite{bib10} and a classical solver named Gurobi for benchmarking. We have considered all the possible number of trades/currencies which can be taken from the given 14 currencies, which range from 2 to 10. From 11 onwards it is not feasible to generate the arbitrage from the given data.

\begin{figure}[!h]
\centering
\begin{subfigure}{0.27\textwidth}
    \centering
        \includegraphics[scale=0.5]{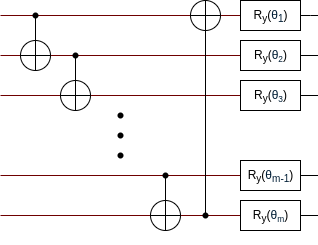}
        \caption{Circuit 1}
        \label{fig:circuit_1}
    \end{subfigure}%
    \begin{subfigure}{0.22\textwidth}
    \centering
        \includegraphics[scale=0.4]{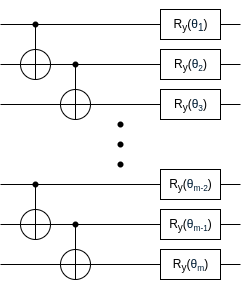}
        \caption{Circuit 2}
        \label{fig:circuit_2}
    \end{subfigure}
\caption{Different Ansatz used in ACE-LS}
\label{fig:ansatz}
\end{figure}

\begin{figure*}[h]
\centering
\includegraphics[scale=0.56]{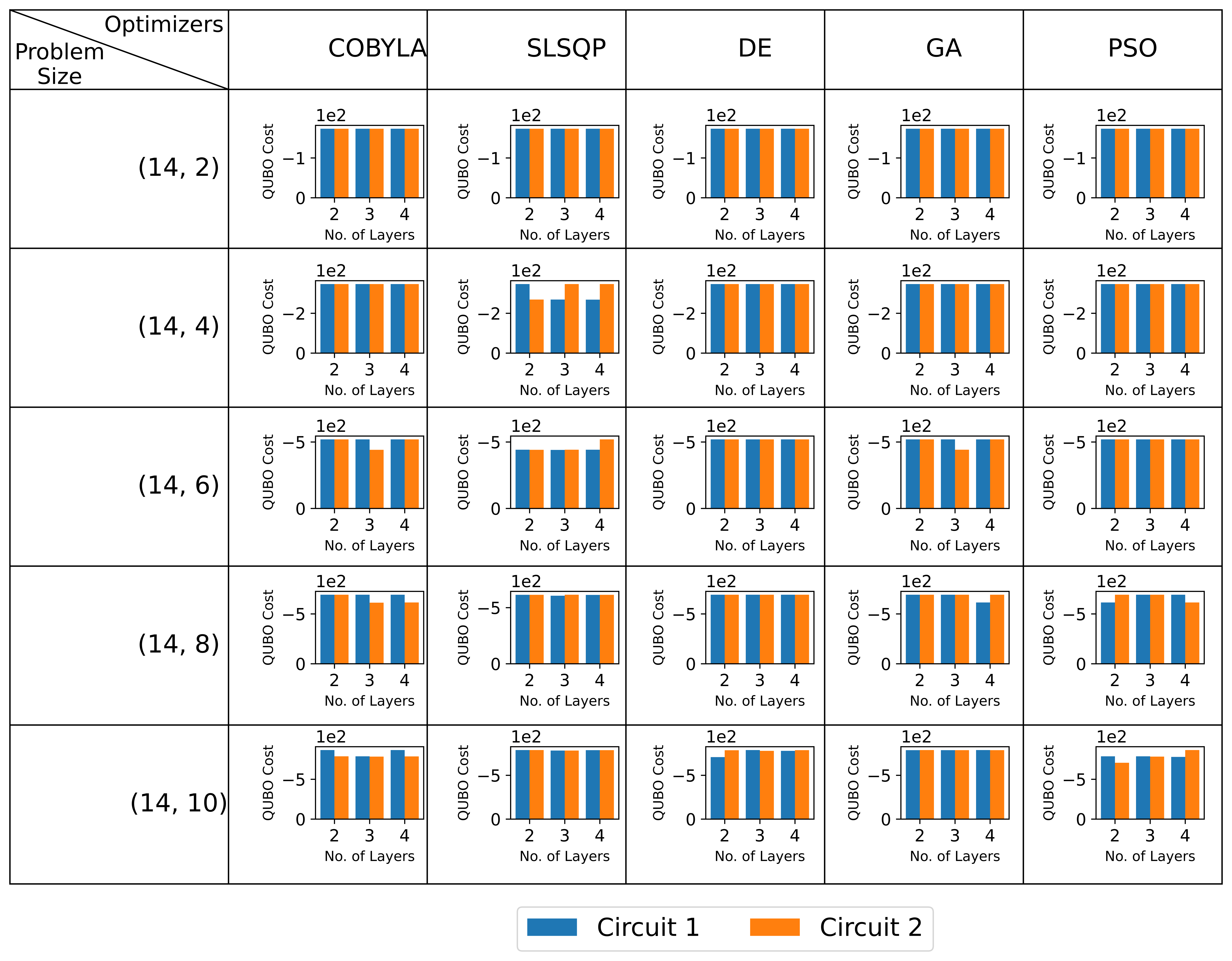}
\caption{Experiments with different circuit, layers and  optimizers in ACE-LS approach. Circuit 2 with two layers, and by using DE optimizer, it gives lowest cost for all problem sizes.}
\label{fig:experiment}
\end{figure*}

In Figure \ref{fig:experiment}, we have shown ACE performance with different circuits, layers and classical optimizers which are used to train the circuit parameters. Circuit 1 and Circuit 2 have different ansatz as shown in Figure \ref{fig:ansatz}, with different classical optimizers that includes local optimizers (e.g., COBYLA and SLSQP), evolutionary optimizer (e.g., Differential Evolution (DE)) and heuristic optimizers (e.g., Genetic Algorithm (GA) and Particle Swarm Optimization (PSO)) respectively. The circuit depth is constant with respect to qubit counts in Circuit 2 compared to Circuit 1 and the number of qubits is provided with respect to number of variables in Table \ref{tab:profit} with the formula $\left\lceil log_2(n)\right\rceil+1$, where $n$ is number of variables in the QUBO. In our experiment, we ran all the circuits five times and took the solution that gives best objective value of the QUBO, and we observed that DE performs quite well compared to other optimizers in all problem sizes. Based on our experiments in ACE-LS, we have used Circuit 2 with two layers, and DE as classical optimizer to find the optimal arbitrage opportunity. 

In Table \ref{tab:profit}, the arbitrage profit has been calculated by the products of the currency exchange rates in the generated cycle that is $[(\prod_{k=1}^{K-1} w_{k,k+1})w_{K, 1}-1]$, where $k$ is the position in the cycle and $w_{k,k+1}$ is the exchange rate from currency at position $k$ to next position currency denoted by $(k+1)$. The profit values have been calculated in pip where 1 pip = .0001 difference in price, e.g. 9.45 pip means 0.000945 unit of price difference in the currency, so the user will generate 0.000945 unit currency profit from 1 unit currency in generated arbitrage. In this table, we observe that QA, ACE-LS performs quite well with respect to classical solvers and provides optimal output on small problems which are benchmarked with Gurobi. Also, these approaches provide profit on large problems but not the best one, and the highlighted values show the most profit for each problem size.

\begin{figure}[!h]
\centering
\includegraphics[scale=0.46]{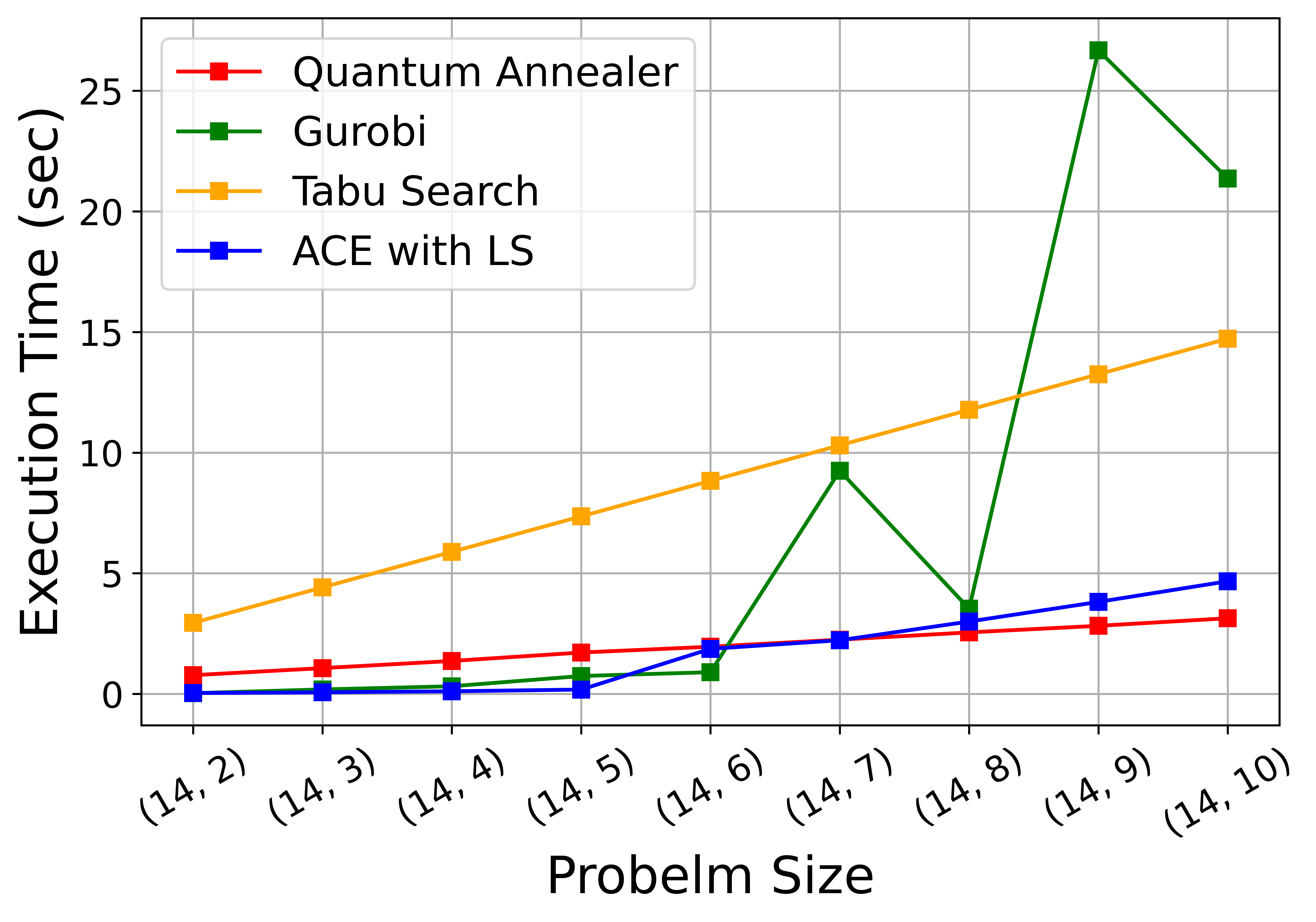}
\caption{Execution Time for Different Solvers}
\label{fig:time}
\end{figure}

Figure \ref{fig:time} shows the execution time of each of the solvers for different problem sizes. We observed QA and ACE-LS take less time than classical solvers. It scales linearly with problem size whereas Gurobi grows exponentially.

\begin{table}[]
\caption{Profit Generated from Different Solvers}
\label{tab:profit}
\begin{tabular}{|c|c|c|c|c|c|c|}
\hline
\begin{tabular}[c]{@{}c@{}}Problem \\ Size\\  (N, K)\end{tabular} & \begin{tabular}[c]{@{}c@{}}No. of \\ variables\\  N(K+1)\end{tabular} & \begin{tabular}[c]{@{}c@{}}No. of \\ Qubits in \\ ACE-LS\end{tabular} & \begin{tabular}[c]{@{}c@{}}QA\\ (pip)\end{tabular} & \begin{tabular}[c]{@{}c@{}}Gurobi\\ (pip)\end{tabular} & \begin{tabular}[c]{@{}c@{}}TS\\ (pip)\end{tabular} & \begin{tabular}[c]{@{}c@{}}ACE \\ - LS\\ (pip)\end{tabular} \\ \hline
(14, 2)                                                           & 42                                                                     & 7                                                                     & \textbf{9.45}                                      & \textbf{9.45}                                          & \textbf{9.45}                                      & \textbf{9.45}                                               \\ \hline
(14, 3)                                                           & 56                                                                     & 7                                                                     & \textbf{9.53}                                      & \textbf{9.53}                                          & \textbf{9.53}                                      & \textbf{9.53}                                               \\ \hline
(14, 4)                                                           & 70                                                                     & 8                                                                     & 8.56                                               & \textbf{8.64}                                          & \textbf{8.64}                                      & \textbf{8.64}                                               \\ \hline
(14, 5)                                                           & 84                                                                     & 8                                                                     & \textbf{17.48}                                     & \textbf{17.48}                                         & \textbf{17.48}                                     & \textbf{17.48}                                              \\ \hline
(14, 6)                                                           & 98                                                                     & 8                                                                     & 12.56                                              & \textbf{16.94}                                         & \textbf{16.94}                                     & \textbf{16.94}                                              \\ \hline
(14, 7)                                                           & 112                                                                    & 8                                                                     & 14.16                                              & \textbf{16.33}                                         & \textbf{16.33}                                     & \textbf{16.33}                                              \\ \hline
(14, 8)                                                           & 126                                                                    & 8                                                                     & 13.74                                              & \textbf{14.75}                                         & \textbf{14.75}                                     & 9.06                                                        \\ \hline
(14, 9)                                                           & 140                                                                    & 9                                                                     & 10.69                                              & \textbf{13.8}                                          & \textbf{13.8}                                      & 3.85                                                        \\ \hline
(14, 10)                                                          & 154                                                                    & 9                                                                     & 11.63                                              & \textbf{11.79}                                         & \textbf{11.79}                                     & 2.08                                                        \\ \hline
\end{tabular}
\end{table}

\begin{figure}
\centering
\includegraphics[scale=0.37]{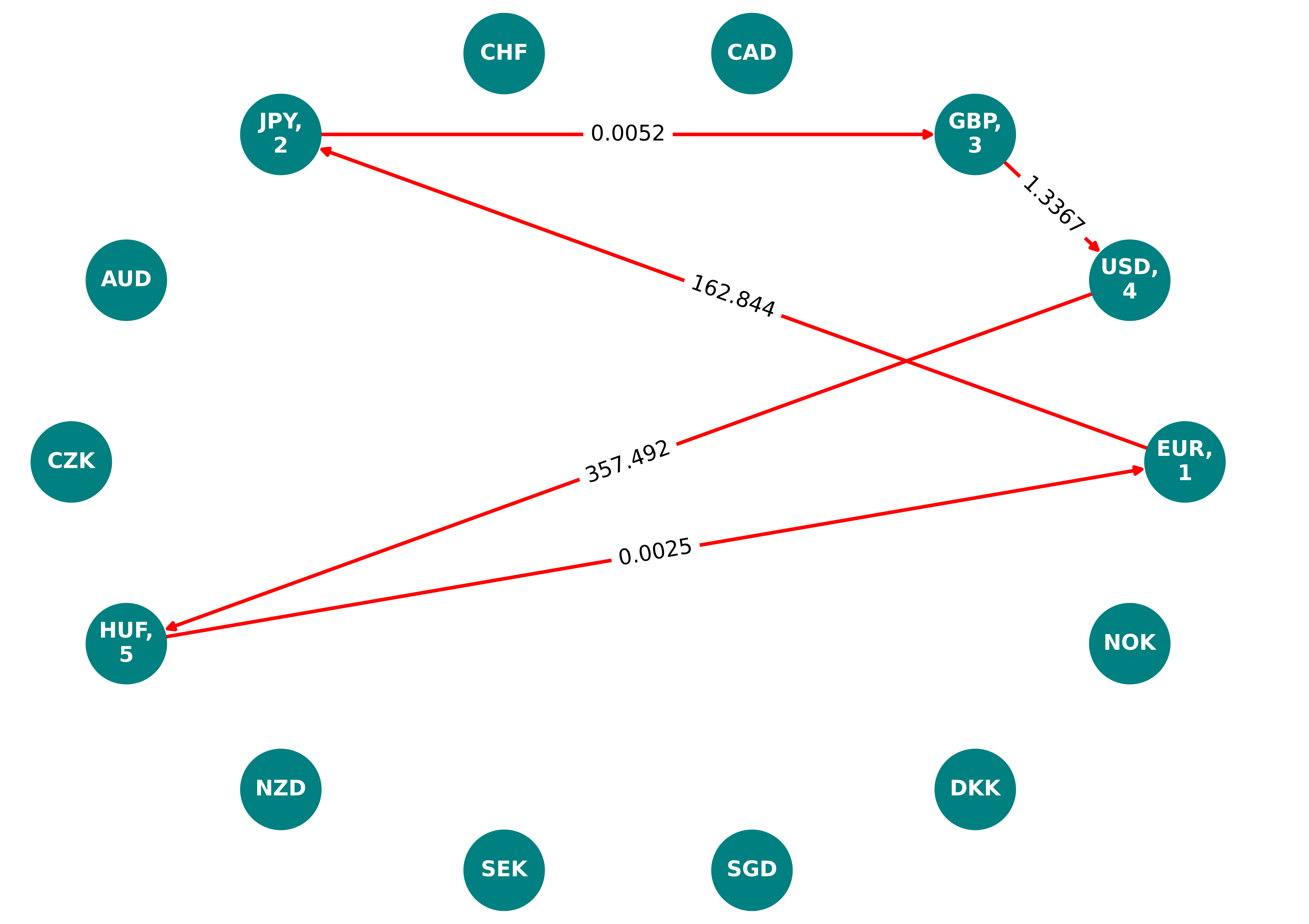}
\caption{Optimal Arbitrage Cycle for Problem Size (14, 5)}
\label{fig:graph}
\end{figure}

Figure \ref{fig:graph} shows the optimal arbitrage cycle generated from the problem size (14, 5), i.e., 5 currency arbitrage cycle starting from EURO (EUR) indicating 1. The sequence of the cycle is denoted by the corresponding number with the currencies and respective currency rates are shown in this figure.

\section{Conclusion and Future Works}
This study highlights the effectiveness of quantum and hybrid quantum-classical methods in addressing real-time currency arbitrage, utilizing price discrepancies across 14 currency pairs from \cite{bib13}. The proposed enhanced mathematical model, with simple-cycle preservation constraints, ensures valid trading cycles and removes infeasible substructures, offering a robust optimization framework. Comparative evaluation of QA, ACE-LS, Gurobi, and Tabu Search, show that QA, ACE-LS outperform Gurobi and TS with respect to execution time, scaling linearly with problem size compared to Gurobi's exponential growth. The solution quality of QA and ACE-LS are comparable with the benchmarks set by Gurobi and TS, affirming the promise of quantum approaches. In currency arbitrage execution time is critical, and QA and ACE-LS can be considered as an initial step toward an industry-level quantum solution. In contrast, real-time currency arbitrage demands execution within microseconds to milliseconds, achievable with HPC systems using parallel computing across CPUs, GPUs, and often FPGAs with ultra-low latency infrastructure. 

Future research will focus on executing ACE-LS on real quantum hardware to validate its performance on a practical level. Investigating advanced hybrid classical-quantum algorithms to further reduce the execution times and improve the solution quality. Also, we can explore the LS approach in quantum domain, so that the classical overhead can be reduced for improving solution quality. In future, the proposed approach for currency arbitrage can be solved on neutral atom-based quantum computers, leveraging their scalability and long coherence times. These advancements aim to enhance computational efficiency and practical applicability in real-world financial scenarios.


\bibliographystyle{unsrt}
\bibliography{reference}

\begin{thebibliography}{10}

\bibitem{bib14}
Zhenyu Cui, Wenhan Qian, Stephen Taylor, and Lingjiong Zhu.
\newblock Detecting and identifying arbitrage in the spot foreign exchange
  market.
\newblock {\em Quantitative Finance}, 20(1):119--132, 2020.

\bibitem{bib1}
Richard Bellman.
\newblock On a routing problem.
\newblock {\em Quarterly of applied mathematics}, 16:87 -- 90, 1958.

\bibitem{bib2}
Zhenyu Cui, Wenhan Qian, Stephen Taylor, and Lingjiong Zhu.
\newblock Detecting and identifying arbitrage in the spot foreign exchange
  market.
\newblock {\em Quantitative Finance}, 20(1):119--132, 2020.

\bibitem{bib3}
Rachel Smith.
\newblock A discussion of linear programming and its application to currency
  arbitrage detection.
\newblock 2013.

\bibitem{bib4}
Gilli Rosenberg.
\newblock Finding optimal arbitrage opportunities using a quantum annealer.
\newblock {\em 1QB Information Technologies Write Paper}, pages 1--7, 2016.

\bibitem{bib5}
Esteban Aguilera, Jins de~Jong, Frank Phillipson, Skander Taamallah, and Mischa
  Vos.
\newblock Multi-objective portfolio optimization using a quantum annealer.
\newblock {\em Mathematics}, 12(9):1291, 2024.

\bibitem{bib6}
Shengbin Wang, Peng Wang, Guihui Li, Shubin Zhao, Dongyi Zhao, Jing Wang, Yuan
  Fang, Menghan Dou, Yongjian Gu, Yu-Chun Wu, et~al.
\newblock Variational quantum eigensolver with linear depth problem-inspired
  ansatz for solving portfolio optimization in finance.
\newblock {\em Science China Information Sciences}, 68(8):1--11, 2025.

\bibitem{bib7}
Supreeth~Mysore Venkatesh, Antonio Macaluso, Marlon Nuske, Matthias Klusch, and
  Andreas Dengel.
\newblock Qubit-efficient variational quantum algorithms for image
  segmentation.
\newblock {\em 2024 IEEE International Conference on Quantum Computing and
  Engineering (QCE)}, 1:450--456, 2024.

\bibitem{bib8}
Sangram Deshpande, Elin~Ranjan Das, and Frank Mueller.
\newblock Currency arbitrage optimization using quantum annealing, qaoa and
  constraint mapping.
\newblock {\em arXiv preprint arXiv:2502.15742}, 2025.

\bibitem{bib8a}
Mayowa Ayodele.
\newblock Penalty weights in qubo formulations: Permutation problems.
\newblock pages 159--174, 2022.

\bibitem{bib15}
Atanu Rajak, Sei Suzuki, Amit Dutta, and Bikas~K Chakrabarti.
\newblock Quantum annealing: An overview.
\newblock {\em Philosophical Transactions of the Royal Society A},
  381(2241):20210417, 2023.

\bibitem{bib9}
Finley~Alexander Quinton, Per Arne~Sevle Myhr, Mostafa Barani, Pedro Crespo~del
  Granado, and Hongyu Zhang.
\newblock Quantum annealing applications, challenges and limitations for
  optimisation problems compared to classical solvers.
\newblock {\em Scientific Reports}, 15(1):12733, 2025.

\bibitem{bib13}
{OANDA}.
\newblock {OANDA Currency Converter}, 2025.

\bibitem{bib11}
Ville Bergholm, Josh Izaac, Maria Schuld, Christian Gogolin, Shahnawaz Ahmed,
  Vishnu Ajith, M~Sohaib Alam, Guillermo Alonso-Linaje, B~AkashNarayanan, Ali
  Asadi, et~al.
\newblock Pennylane: Automatic differentiation of hybrid quantum-classical
  computations.
\newblock {\em arXiv preprint arXiv:1811.04968}, 2018.

\bibitem{bib12}
{D-Wave Documentation}.
\newblock {D-Wave Ocean}, 2018.

\bibitem{bib10}
Fred Glover.
\newblock Tabu search: A tutorial.
\newblock {\em Interfaces}, 20(4):74--94, 1990.

\end{thebibliography}

\end{document}